\documentclass[10pt]{article}
\usepackage[dvips]{graphicx}
\usepackage{amssymb}
\usepackage{amsmath}
\usepackage{amsfonts}

\def\one{1\hskip -1mm{\rm l}}
\begin{document}

\begin{center}
{\Large \bf \sf
      A New Boson  realization of Fusion Polynomial Algebras in Non-Hermitian Quantum Mechanics : $\gamma$-deformed $su(2)$ generators, Partial $\mathcal{PT}$-symmetry and Higgs algebra }

\bigskip\vspace{1cm}
{\sf Arindam Chakraborty$^1$\footnote{e-mail: arindam.chakraborty@heritageit.edu}},

\bigskip

{\em $^1$Department of Physics, Heritage Institute of Technology, Kolkata-700107, India}

\end{center}

\vskip 20pt
\begin{center}
{\bf Abstract}
\end{center}

\par A $\gamma$-deformed version of $su(2)$ algebra with non-hermitian generators has been obtained from a bi-orthogonal system of vectors in $\bf{C^2}$. The related Jordan-Schwinger(J-S) map is combined with boson algebras to obtain a hierarchy of fusion polynomial algebras. This makes possible the construction of Higgs algebra of cubic polynomial type. Finally the notion of partial $\mathcal{PT}$ symmetry has been introduced as characteristic feature of some operators as well as their eigenfunctions. The possibility of  partial $\mathcal{PT}$-symmetry breaking is also discussed. The deformation parameter $\gamma$ plays a crucial role in the entire formulation and non-trivially modifies the eigenfunctions under consideration.


\section{Introduction}
The study of quantum mechanical formalism involving non-hermitian operators has gained much attention for last couple of decades engendering a major epistemological break in the practice of conventional quantum mechanics in a typically hermitian framework\cite{morettibook}. Since the identification of non-hermitian operators with real spectrum and Bender and Boettcher's\cite{bender98, bender02} attribution of this possibility to space-time reflection symmetry( $\mathcal{PT}$-symmetry) so many studies have been undertaken in a variety of contexts relating to discrete symmetries\cite{brody16,  mosta02}. Non-hermitian formulation is often found to be meaningful when complications arise in the hermitian formalism(cases with complex potentials), in the study of so called resonance phenomena  associated with nuclear, atomic or molecular systems or even with nano-structured materials or condensates, in understanding systems which are not so quantum mechanical in sense but their physical behaviour is quite amenable to quantum language(for example classical statistical mechanical systems, biological systems with diffusion, light propagation in wave guides) and many other fields where even the conventional quantum mechanics has already shown success\cite{nimrodbook}.

\par The present article stems from a few of recent studies done by Brody\cite{brody16, brody14}regarding the construction of non-hermitian $su(2)$ generators with real eigenvalues. It has been claimed\cite{brody14} that a parallel formalism of quantum mechanics, at least in the framework of finite dimensional Hilbert space, is possible even for relaxing the requirement of hermiticity of the observables and unitary evolution of the physical system under consideration. The demand of orthogonality has been replaced by the notion of bi-orthogonality of eigenvectors discerned from a non-hermitian operator and its hermitian conjugate. In our case a bi-orthogonal  system has been formulated starting from a pair of orthonormal bases and subsequent construction of Riesz bases\cite{nonselfbook} with the help of a suitably chosen transformation. Such a bi-orthogonal system of vectors can be used to construct a set of generators of a Lie algebra which in the present setting comes out as  a parametric deformation of $su(2)$ algebra. The related Jordan-Schwinger operators $\{J_0^{\gamma}, J_{\pm}^{\gamma}\}$ have been constructed where the operator $J_0^{\gamma}$ becomes a non-hermitian operator.

\par It is a well known fact that there exist seemingly different physical systems which can have identical algebraic structure (algebraic isomorphism) that makes them quite akin to each other in terms of symmetries, conservation principles and many other attributes. In order to appreciate a number of Lie algebraic aspects resulting out of a two-boson realization of a newly obtained deformed algebra it is customary to construct the so called Jordan-Schwinger map\cite{biedenbook, kim87, frad94} which is a two boson realization of Lie algebra. Such realization has been used extensively in the study of atomic\cite{condon80}, nuclear\cite{klein91} and molecular structures\cite{halo83, iachello95}. It is interesting to note that the complexification of the deformed algebra which otherwise resembles $su(2)$ for $\gamma=0$ is not spectrum generating in general for all non-zero values of $\gamma$. However a $\gamma$ deformed version of complexification is possible at the level of Jordan-Schwinger realization which is spectrum generating in character and  resembles the corresponding realization of $su(2)$ for $\gamma=0$. The reality of the spectrum corresponds to $\gamma\leq 1$, a fact which is closely related to property of bi-orthogonality. The possibility of existence of a class of polynomial fusion algebras\cite{sunil01, sunil02, ruan01} has been explored by combining the J-S maps of either two deformed algebras or one such deformed algebra and another boson algebra. While the first attempt directly leads to cubic algebra, the second one   results to a hierarchy of polynomial algebras. It is interesting to note that each level of hierarchy is equipped with a new $\gamma$-deformed parameter $\{\omega_{n-1}:n\geq 2\}$ that can be obtained recursively and plays crucial role in the formulation of respective Casimirs\cite{sunil02}. Cubic algebra, on the other hand produces the so called Higgs algebra under certain condition involving the deformation parameter $\gamma$. Higgs algebra is one of the earliest candidates of Polynomial Angular Momentum Algebra(PAMA)\cite{ruan06,sunil02}. It has been introduced\cite{higgs79} to establish the existence of hidden symmetry for coulomb and oscillator potential in a space of constant curvature also understood as a second order approximation of $su_q(2)$\cite{zhed92}. In the present occasion two such isomorphic multi-boson realizations\cite{ruan06} of Higgs algebra have been constructed. The same is obtained from the fusion of algebras $su_{\gamma}(2)$ and $su_{\gamma}(1, 1)$. In all of the aforementioned cases the deformation parameter $\gamma$ plays a crucial role in this construction. The related commutation relations result to multi-boson Hamiltonians involving central elements of the algebra, all of which are non-hermitian operators. Starting from Higgs algebra a deformed version of Hahn algebra\cite{zhed92} is also obtained.

\par The presence of $\mathcal{PT}$-symmetry in non-hermitian Hamiltonian\cite{garcia17} has been widely exploited  in describing various systems and processes especially in open regimes with balanced gain and loss like laser absorbers\cite{longhi10}, ultracold threshold phonon lasers\cite{jing14}, defect states in special beam dynamics in spatial lattices\cite{rogen13}. Furthermore the relevance of $\mathcal{PT}$-symmetry has been appreciated in the behaviour of quantum circuit based on nuclear magnetic resonance\cite{zheng13}, microwave cavities\cite{bittner12}, super-conductivity\cite{rubin07} and Bose-Einstein condensates\cite{krei16}. At the theoretical level the behaviour of $\mathcal{PT}$-symmetric operators along with the nature of their target vector space\cite{brody17} and their significance  in pseudo-hermitian quantum mechanics\cite{mosta05} have also become some interesting areas of investigation.

\par In a recent article by Beygi et. al.\cite{beygi15} the issue of partial $\mathcal{PT}$ symmetry has been investigated for $N$-coupled harmonic oscillator Hamiltonian with purely imaginary coupling term whereas the reality and partial reality of the spectrum are claimed to have direct correspondences with the classical trajectory. The interpretation of all such symmetries can be understood both at the level of Hamiltonian as well as in its eigenstates\cite{bender02}. Here we have introduced the notion of partial $\mathcal{PT}$ symmetry in relation to boson operators which eventually helps us to understand the presence of the same symmetry in many other operators obtained earlier. A Bargmann-Fock type correspondence relating to the Boson operators has been set up.These operators leaves invariant the  space of homogeneous polynomial involving two indeterminates. This consideration  leads to  a tri-diagonal representation of the operator $J^{\gamma}_0$. The behaviour of its eigenvalues is compared with that of the hermitian case in view of Gershgorin disk theorem\cite{friedbook, eigenbook}. Finally the eigenvalues are obtained explicitly as functions of the deformation parameter $\gamma$ via an iterative algorithm of eigen-decompsition\cite{sandry13}. The eigenvalues are found to be real and eigenvectors are modified in a non-trivial way. Partial parity symmetry has also been investigated in eigenstates in the relevant representation space\cite{ruan01, ruan06}. It is readily observed that the presence of non-hermiticity in the aforementioned operator amounts to sacrificing its own global $\mathcal{PT}$-symmetry and also that in some of their eigenstates. In some cases the breaking of such partial $\mathcal{PT}$-symmetry is also noticed. This seems to be one of the generic features of any J-S realization in the non-hermitian quantum mechanics.

\section{Auerbach Biorthogonal System and Deformed $su(2)$ Algebra}
Given two orthogonal vectors $\{\vert u_j\rangle=\left(\begin{array}{c}
  1  \\
  (-1)^{j-1} 
    \end{array} \right):j=1, 2\}$, $\langle u_j \vert u_k\rangle=\delta_{jk}$ one can construct Pauli spin matrices (as conventional generators of $su(2)$ algebra) like the following
\begin{eqnarray}
\sigma_m=\frac{i^{m+1}}{2}c^{(m)}_{jk}\vert u_j \rangle\langle u_k\vert\:\: : m=1, 2, 3
\end{eqnarray}
where $c^{(1)}_{jk}=(-1)^j\delta_{jk}$ and $c^{(3)}_{jk}=(-1)^jc^{(2)}_{jk}=(1-(-1)^jc^{(1)}_{jk})$.
This leads to $su(2)$ commutation relation $[\sigma_l, \sigma_m]=i\epsilon_{lmn}\sigma_n$. 
\par Now considering Riesz bases $\{\vert \chi_j\rangle=T\vert u_j\rangle\::j=1, 2\}$ with $T=\cos\frac{\theta}{2}\one_2+2\cos\frac{1}{2}\phi \sin\frac{\theta}{2}\sigma_1-2\sin\frac{1}{2}\phi \sin\frac{\theta}{2}\sigma_2$ we get  $\{\vert \phi_j\rangle=\omega_0(T^{-1})^{\dagger}\vert u_j\rangle\::j=1, 2\}$ where $\omega_0=\cos \theta$.  As $\langle \phi_j\vert\chi_k\rangle=0\: \forall j\neq k$, $\{\vert \chi_j\rangle; \vert \phi_j\rangle\}$ represents a bi-orthogonal system of vectors. A theorem due to Auerbach\cite{auerbook} ensures that for a finite dimensional Banach space(here $C^2$) such bi-orthogonal system of vectors is always available. One can therefore construct a $\gamma$-deformed algebra $su_{\gamma}(2)$  with the following basis.
\begin{eqnarray}
\sigma_m^{\gamma}=\frac{i^{m+1}}{2}\frac{c^{(m)}_{jk}}{\omega_0^{\delta_{m2}}}\vert \phi_j \rangle\langle \chi_k\vert\:\: : m=1, 2, 3
\end{eqnarray}
where we have taken $\phi=\pi$ and $\gamma=\sqrt{1-\omega_0^2}=\sin\theta, \:\:-\pi/2<\theta<\pi/2$.
\par  The set of matrices $\{\sigma^{\gamma}_1,\sigma^{\gamma}_2,\sigma^{\gamma}_3\}$ provides the following commutation relation
\begin{equation}
[\sigma^{\gamma}_l, \sigma^{\gamma}_m]=i\epsilon_{lmn}(1-\gamma^2\delta_{k2})\sigma^{\gamma}_n
\end{equation}
This algebra is said to be a deformation of $su(2)$ in the sense that apart from tracelessness all most all attributes of $su(2)$ are sacrificed due to this deformation but all of them can be recovered by letting $\gamma=0$. The generators $\{\sigma_m^{\gamma}:m=1, 3\}$ are non-hermitian in the conventional sense of inner product in $\bf{C^2}$.

\section{Jordan-Schwinger Map}
Considering a set of boson operators $\{a_{\alpha}, a_{\alpha}^{\dagger}\:\:\alpha=1,2\}$ with $[a_{\alpha}, a^{\dagger}_{\beta}]=\delta_{\alpha\beta}$ one can define the operator-valued Jordon-Schwinger(henceforth J-S) realization\cite{biedenbook} (a lie algebra isomorphism) of $su(2)$ with the bilinear map $\{J_m=(\sigma_m)_{\mu\nu}a^{\dagger}_{\mu}a_{\nu}\::m=1, 2, 3\}$. This
leads to
\begin{eqnarray}
J_m=\frac{i^{m-1}}{2}c_{jk}^{(4-m)}a^{\dagger}_ja_k\:\: :m=1, 2, 3
\end{eqnarray}
Defining $J_{\pm}=J_1\pm iJ_2$ and using $[a_{\alpha}^{\dagger}a_{\beta}, a_{\mu}^{\dagger}a_{\nu} ]=a_{\alpha}^{\dagger}a_{\nu}\delta_{\beta\mu}- a_{\mu}^{\dagger}a_{\beta}\delta_{\alpha\nu}$ we obtain a spectrum generating algebra(SGA)(taking $J_3=J_0$):
\begin{eqnarray}
[J_0, J_{\pm}]&=&\pm J_{\pm}\nonumber\\
{[J_+, J_- ]}&=&2J_0
\end{eqnarray}
The above realization can be viewed as an occupation number representation of the two dimensional isotropic harmonic oscillator. The Casimir of this algebra is obviously given by $\mathcal{C_J}=J_0(J_0\pm1)+J_{\mp}J_{\pm}$. On the other hand the J-S realization of $su_{\gamma}(2)$-algebra does not immediately produce a spectrum generating algebra in the same way as opposed to the usual one. In the $\gamma$-deformed case J-S map gives
 \begin{eqnarray}
J_m^{\gamma}=\frac{i^{m-1}}{2}\left[c_{jk}^{(4-m)}+i^m(1-\delta_{m2})\gamma c_{jk}^{(m)}\right]a^{\dagger}_ja_k\:\::m=1, 2, 3
\end{eqnarray} 
However introducing a pair of new ladder operators with exponent $p$,  $J_{\pm}^{\gamma}=\omega_0^{-p}J_1^{\gamma}\pm i\omega_0^{-p+1}J_2^{\gamma}$ where $\omega_0=(1-\gamma^2)^{1/2}$ and considering $J^{\gamma}_{3}=J^{\gamma}_0$ it is possible to construct  the following spectrum generating algebra (with $J_0^{\gamma}$ being non-hermitian)
\begin{eqnarray}
[J^{\gamma}_0, J_{\pm}^{\gamma}]&=&\pm \omega_0(\gamma) J^{\gamma}_{\pm}\nonumber\\
{[J^{\gamma}_+, J^{\gamma}_- ]}&=&2\omega_0^{-2p+1}(\gamma)J^{\gamma}_0
\end{eqnarray}
Correspondingly the Killing metric tensor becomes
\begin{eqnarray}
 g_{\mu\nu}=2\left(\begin{array}{ccc}
   \omega_0^2 &  0 & 0\\
   0 & 0 & 2\omega_0^{-2p+2}\\
   0 & 2\omega_0^{-2p+2} & 0 
    \end{array} \right)
\end{eqnarray}
and hence the Casimir becomes $\mathcal{C_J^{\gamma}}=\frac{1}{\omega_0^2}J_0^{\gamma}(J_0^{\gamma}\pm\omega_0)+\omega_0^{2p-2}J_{\mp}^{\gamma}J_{\pm}^{\gamma}$. It is interesting to note that for $\gamma=0$ the above results correspond to those of $su(2)$ algebra. The algebra is compact and semi-simple as the Killing from is negative definite for all allowed values of $\omega_0$ or $\gamma$. 

\par On similar ground a deformed $su(1, 1)$ algebra ($su_{\mu}(1, 1)$) can be conceived by defining $\{\tau^{\mu}_m=i^{1+3\delta_{m3}}\sigma^{\mu}_m\::m=1, 2, 3\}$ as basis with the commutation $[\tau^{\mu}_k, \tau^{\mu}_l]=i^{1+2\delta_{m3}}\epsilon_{klm}(1-\mu^2\delta_{m2})\tau^{\mu}_m$. The corresponding J-S is constructed as $Z_m=W^{\sharp}\tau_m W$. Here, $W=(b_1\:\:b_2^{\dagger})$ and $W^{\sharp}=(b_1^{\dagger}\:\:-b_2)$. Defining the corresponding ladder operators $Z_{\pm}=\omega_0^{-q}Z_1\pm i\omega_0^{-q+1}Z_2$ following commutations result

\begin{eqnarray}
[Z^{\mu}_0, Z_{\pm}^{\mu}]&=&\pm \omega_0(\mu) Z^{\mu}_{\pm}\nonumber\\
{[Z^{\mu}_+, Z^{\mu}_- ]}&=&-2\omega_0^{-2q+1}(\mu)Z^{\mu}_0
\end{eqnarray}
 Casimir of this algebra is given by $\mathcal{C_Z^{\mu}}=Z^{\mu}_{\pm}Z^{\mu}_{\mp}-Z_0^{\mu}(Z_0^{\mu}\pm\omega_0)$.

\section{Quadratic and Cubic algebras}
One can view the above mentioned algebra as a polynomial algebra(equation-7) in the following way
\begin{eqnarray}
[J^{\gamma}_0, J_{\pm}^{\gamma}]&=&\pm \omega_0 J^{\gamma}_{\pm}\nonumber\\
{[J^{\gamma}_+, J^{\gamma}_- ]}&=& P^{(1)}(J_0^{\gamma}, \omega_0)
\end{eqnarray}
Identifying $g^{(2)}(J_0^{\gamma}; \omega_0)=(J_0^{\gamma})^2+{\omega_0}J_0^{\gamma}$ as a second order polynomial in $J_0^{\gamma}$, it is readily observed $P^{(1)}=\frac{1}{\omega_0^{2p}}[g^{(2)}(J_0^{\gamma}; \omega_0)-g^{(2)}(J_0^{\gamma}; -\omega_0)]$ and the Casimir has the expression $\frac{\mathcal{C_J^{\gamma}}}{\omega_0^{2p-2}}=J^{\gamma}_{\mp}J^{\gamma}_{\pm}+\frac{1}{\omega_0^{2p}}g^{(2)}(J_0^{\gamma}; \pm\omega_0))$.

\subsection{Fusion of   $su_{\gamma}(2)$ and boson algebras}
Now fusion of a given polynomial algebra and a boson algebra can produce higher order polynomial algebras.  An $n$-th order $(n\geq 2)$ polynomial algebra  involving the generators $\{S_0^{\gamma},S_{\pm}^{\gamma}\}$can be defined in terms of following commutation relations
\begin{eqnarray}
[S^{\gamma}_0, S_{\pm}^{\gamma}]&=&\pm \omega_{n-1} S^{\gamma}_{\pm}\nonumber\\
{[S^{\gamma}_+, S^{\gamma}_- ]}&=& s_{n-2}^2P^{(n)}(S_0^{\gamma}; \omega_0, \omega_1,\dots, \omega_{n-2})
\end{eqnarray}
admitting a Casimir
\begin{equation}
\mathcal{C_S^{\gamma}}=S^{\gamma}_{-}S^{\gamma}_{+}+ s_{n-2}^2\omega_{n-1}g^{(n+1)}(S_0^{\gamma}, \omega_0, \omega_{1}, \dots, \omega_{n-1})=S^{\gamma}_{+}S^{\gamma}_{-}+ s_{n-2}^2\omega_{n-1}g^{(n+1)}(S_0^{\gamma}, \omega_0, \omega_{1}, \dots, -\omega_{n-1})
\end{equation}
where $g^{(n+1)}$ is a polynomial of order $n+1$ and $\omega_{n}$ follows the recursion $\omega_n=\frac{1}{2}(1+\omega_{n-1})\:\::n=1, 2, 3\dots$

\par A $\gamma$-deformed quadratic algebra ($n=2$) can be obtained as a fusion algebra as follows. 
\par Let us take a boson algebra with generators $a_3,a_3^{\dagger}$ with $ [a_3,a_3^{\dagger}]=1$ and define
\begin{eqnarray}
R_0^{\gamma}&=&\frac{1}{2}(J_0^{\gamma}-M_0)\nonumber\\
\Lambda_0^{\gamma}&=&\frac{1}{2}(J_0^{\gamma}+M_0)\nonumber\\
R_+^{\gamma}&=&s_0J_+^{\gamma}a_3\nonumber\\
R_-^{\gamma}&=&s_0J_-^{\gamma}a_3^{\dagger}
\end{eqnarray}
where $M_0=a_3^{\dagger}a_3$
The resulting quadratic algebra satisfies the following commutation relation
\begin{eqnarray}
[R^{\gamma}_0, R^{\gamma}_{\pm}]&=&\pm \omega_1(\gamma)R^{\gamma}_{\pm}\nonumber\\
{[R^{\gamma}_+, R^{\gamma}_- ]}&=&s_0^2P^{(2)}(R_0^{\gamma}; \omega_0)
\end{eqnarray}
where $P^{(2)}(R_0^{\gamma}; \omega_0)=[A_1(\omega_0)(R^{\gamma}_0)^2+B_1(\omega_0)R^{\gamma}_0+C_1(\omega_0)]$ along with the coefficients as rational functions of $\omega_0$ as follows.
\begin{eqnarray}
A_1(\omega_0)&=&-\frac{1+2\omega_0}{\omega^{2p}_0}\nonumber\\
B_1(\omega_0)&=&-\frac{2\Lambda_0^{\gamma}-\omega_0}{\omega^{2p}_0}\nonumber\\
C_1(\omega_0)&=&\frac{(-1+2\omega_0)(\Lambda_0^{\gamma})^2+\omega_0\Lambda_0^{\gamma}+\omega_0^2C_J^{\gamma}}{\omega_0^{2p}}
\end{eqnarray}
The Casimir of this algebra can be obtained as
\begin{equation}
\mathcal{C_R^{\gamma}}=R^{\gamma}_-R^{\gamma}_++ s_{0}^2\omega_1g^{(3)}(R^{\gamma}_0,\omega_0, \omega_1)=R^{\gamma}_+R^{\gamma}_-+ s_{0}^2\omega_1g^{(3)}(R^{\gamma}_0,\omega_0, -\omega_1)
\end{equation}
where
\begin{equation}
g^{(3)}(R^{\gamma}_0,\omega_0, \omega_1)=\frac{A_1}{3\omega_1^2}(R_0^{\gamma})^3+\left(\frac{A_1\omega_1+B_1}{2\omega_1^2}\right)(R_0^{\gamma})^2+\left(\frac{A_1\omega_1^2+3B\omega_1+C_1}{6\omega_1^2}\right)R^{\gamma}_0+\frac{C_1}{2\omega_1}
\end{equation}
Similarly a $\gamma$-deformed cubic algebra $(n=3)$ is possible combining the above quadratic algebra and another boson algebra with generators $a_4,a_4^{\dagger}:[a_4,a_4^{\dagger}]=1$. Considering $N_0=a_4^{\dagger} a_4, Q_0^{\gamma}=\frac{1}{2}(R_0^{\gamma}-N_0), \Delta_0^{\gamma}=\frac{1}{2}(R_0^{\gamma}+N_0), Q^{\gamma}_{+}=s_1R_{+}^{\gamma}a_4$ and $Q^{\gamma}_{-}=s_1R_{-}^{\gamma}a_4^{\dagger}$ following relations can be obtained as a four boson realization
\begin{eqnarray}
[Q^{\gamma}_0, Q^{\gamma}_{\pm}]&=&\pm \omega_2(\gamma)Q^{\gamma}_{\pm}\nonumber\\
{[Q^{\gamma}_+, Q^{\gamma}_- ]}&=& s_1^2P^{(3)}(Q_0^{\gamma}; \omega_0, \omega_1)
\end{eqnarray}

where $P^{(3)}(Q_0^{\gamma}; \omega_0, \omega_1)= A_2(\omega_0,\omega_1)(Q^{\gamma}_0)^3+B_2(\omega_0,\omega_1)(Q^{\gamma}_0)^2+C_2(\omega_0,\omega_1)Q_0^{\gamma}+D_2(\omega_0, \omega_1)$
\begin{eqnarray}
A_2&=&-(1+(3\omega_1)^{-1})A_1\nonumber\\
B_2&=&\{1/2-(1+\omega_1^{-1})\Delta_0^{\gamma}\}A_1-(1+(2\omega_1)^{-1})B_1\nonumber\\
C_2&=&\left[(\Delta_0^{\gamma})^2(1-\omega_1^{-1})+\Delta_0^{\gamma}-\frac{\omega_1}{6}\right]A_1+(1/2-\Delta_0^{\gamma}\omega_1^{-1})B_1-(1+\omega_1^{-1})C_1\nonumber\\
D_2&=&\left[(\Delta_0^{\gamma})^3(1-(3\omega_1)^{-1})+\frac{1}{2}(\Delta_0^{\gamma})^2-\frac{\omega_1\Delta_0^{\gamma}}{6}\right]A_1+\left[(\Delta_0^{\gamma})^2(1-(2\omega_1)^{-1})+\frac{1}{2}\Delta_0^{\gamma}\right]B_1\nonumber\\
+\left[\Delta_0^{\gamma}(1-\omega_1^{-1})+\frac{1}{2}\right]C_1
+\mathcal{C_R^{\gamma}}
\end{eqnarray}
The Casimir of the Cubic algebra is given by 
\begin{equation}
\mathcal{C_Q^{\gamma}}=Q_-^{\gamma}Q_+^{\gamma}+ s_{1}^2\omega_2g^{(4)}(Q_0^{\gamma}; \omega_0, \omega_1, \omega_2)=Q_-^{\gamma}Q_+^{\gamma}+ s_{1}^2\omega_2g^{(4)}(Q_0^{\gamma}; \omega_0, \omega_1, -\omega_2)
\end{equation}
 where 
\begin{equation}
g^{(4)}(Q_0^{\gamma}; \omega_0, \omega_1, \omega_2)=\frac{A_2}{4\omega_2^2}(Q_0^{\gamma})^4+\frac{2B_2+3A_2\omega_2^2}{6\omega_2^2}(Q_0^{\gamma})^3+\frac{2C_2+A_2\omega_2^2+2B_2\omega_2}{4\omega_2^2}(Q_0^{\gamma})^2+\frac{6D_2+B_2\omega_2^2+3C_2\omega_2}{6\omega_2^2}Q_0^{\gamma}+\frac{D_2}{2\omega_2}
\end{equation}

It is to be noted that in each generation of fusion polynomial algebra a new $\gamma$ dependent parameter is introduced and this term plays a crucial role in the allied expressions of the polynomial part as well as the Casimir. Similar exercise can be repeated to formulate fusion algebras out of $su_{\gamma}(1, 1)$ and a boson algebra.  In the following we will be interested about a special kind of cubic algebra where the coefficients of quadratic term and the constant part  in eqn-18 type expression are non-existent i. e.; $B_2=0=D_2$. Such type of cubic algebra is known as Higgs algebra.

\subsection{Fusion of two deformed su(2) algebras and Construction of Higgs algebra}
Considering a $\gamma$-deformed algebra $\{J^{\gamma}_{0,\pm}\}$ corresponding to boson operators $\{a_1, a_1^{\dagger}; a_2, a_2^{\dagger}\}$ and a $\mu$-deformed algebra $\{K^{\mu}_{0,\pm}\}$ corresponding to $\{a_3, a_3^{\dagger}; a_4, a_4^{\dagger}\}$ one can define the following quantities
\begin{eqnarray}
H_0^{\gamma,\mu}&=&\frac{1}{2}(J_0^{\gamma}-K_0^{\mu})\nonumber\\
H_{\pm}^{\gamma,\mu}&=&s J_{\pm}^{\gamma}K_{\mp}^{\mu}\nonumber\\
L_0^{\gamma,\mu}&=&\frac{1}{2}(J_0^{\gamma}+K_0^{\mu})
\end{eqnarray}
where $K_{\pm}^{\mu}=\omega_0^{-q}(\mu)K_1^{\mu}\pm i \omega_0^{-q+1}(\mu)K_2^{\mu}$.
This leads to the commutation relations:
\begin{eqnarray}
[H_{0}^{\gamma,\mu},H_{\pm}^{\gamma,\mu}]&=&\pm \frac{1}{2}(\omega_0(\gamma)+\omega_0(\mu))H_{\pm}^{\gamma,\mu}\nonumber\\
{[H_{+}^{\gamma,\mu},H_{-}^{\gamma,\mu}]}&=&s^2[\alpha_0+\alpha_1H_0^{\gamma,\mu}+\alpha_2(H_0^{\gamma,\mu})^2+\alpha_3(H_0^{\gamma,\mu})^3]
\end{eqnarray}
where
\begin{eqnarray}
\alpha_0 &=&\frac{2}{\omega_0^{2q-1}(\mu)\omega_0^{2p-1}(\gamma)}\left[{\mathcal{C_K^{\mu}}}{\omega_0(\mu)}-{\mathcal{C_J^{\gamma}}}{\omega_0(\gamma)}\right]L_0^{\gamma, \mu}+2\Omega_{-}(L_0^{\gamma, \mu})^3\nonumber\\
\alpha_1 &=&\frac{2}{\omega_0^{2p-1}(\gamma)\omega_0^{2q-1}(\mu)}\left[{\mathcal{C_K^{\mu}}}{\omega_0(\mu)}+{\mathcal{C_J^{\gamma}}}{\omega_0(\gamma)}\right]+2\Omega_{+}(L_0^{\gamma, \mu})^2\nonumber\\
\alpha_2 &=&-2\Omega_{-}L_0^{\gamma, \mu}\nonumber\\
\alpha_3 &=&-2\Omega_{+}
\end{eqnarray}
The quantity $\Omega_{\pm}=\frac{1}{\omega_0^{2p-1}(\gamma)\omega^{2q-1}_0(\mu)}\left[\frac{1}{\omega_0(\gamma)}\pm\frac{1}{\omega_0(\mu)}\right]$ is very crucial in the following formulation of a deformed Higgs algebra. For $\gamma=\pm\mu$, $\Omega_+=\frac{2}{(\omega_0)^{2(p+q)-1}}$ and $\Omega_-=0$. Considering further $\mathcal{C_J^{\gamma}}=\mathcal{C_K^{\mu}}=\mathcal{C}$ and writing $H_0^{\gamma, \gamma}=H_0^{\gamma}$ and $L_0^{\gamma, \gamma}=L_0^{\gamma}$ 
\begin{eqnarray}
[H_{0}^{\gamma},H_{\pm}^{\gamma}]&=&\pm (\omega_0(\gamma))H_{\pm}^{\gamma}\nonumber\\
{[H_{+}^{\gamma},H_{-}^{\gamma}]}&=& 4s^2H_{0}^{\gamma}(\gamma_1+\gamma_2(H_0^{\gamma})^2)
\end{eqnarray}
where $\gamma_1=\frac{\mathcal{C}}{\omega_0^{2(p+q)-3}}+\frac{(L_0^{\gamma})^2}{\omega_0^{2(p+q)-1}}$ and $\gamma_2=-\frac{1}{\omega_0^{2(p+q)-1}}$. Choosing $s^2=-\lambda_0 \omega_0^r$
\begin{equation}
[H_{+}^{\gamma},H_{-}^{\gamma}]=4\left[-\lambda_0\left(\frac{{\mathcal{C}}}{\omega_0^{2(p+q)-r-3}}+\frac{(L_0^{\gamma})^2}{\omega_0^{2(p+q)-r-1}}\right)H_0^{\gamma}+\frac{\lambda}{\omega_0^{2(p+q)-r-1}}(H_0^{\gamma})^3\right]
\end{equation}
\par Casimir of the above algebra can be given by
\begin{equation}
\mathcal{C_H^{\gamma}}=H_{\mp}^{\gamma}H_{\pm}^{\gamma}-2\lambda\left(\frac{{\mathcal{C}}}{\omega_0^{2(p+q)-r-3}}+\frac{(L_0^{\gamma})^2}{\omega_0^{2(p+q)-r-1}}\right)H_0^{\gamma}(H_0^{\gamma}\pm\omega_0)+\frac{\lambda}{\omega_0^{2(p+q)-r-1}}(H_0^{\gamma})^2(H_0^{\gamma}\pm\omega_0)^2
\end{equation}

The last term in each of the above expressions has been understood as the quantum addition of classical Casimir operator\cite{zhed92}. 

Higgs algebra has been introduced in the context of non-relativistic Kepler problem in spaces with curvature $\kappa$. In the above results $s^2$ can be chosen as $\pm \kappa$ depending upon whether the curvature is negative or positive corresponding to hyperboloid and sphere respectively\cite{higgs79}. The system is described by the Hamiltonian
\begin{equation}
\mathcal{H}_{\kappa}=\frac{1}{2}[\pi_j\pi_j+\kappa A_0^2]-\frac{\mu_0}{r}
\end{equation}
where $\mu$ is a constant number, $A_0$ is a two dimensional rotation operator, $\pi_j(j=1, 2)$, two components of momentum operator which is given by $\pi_j=p_j-\kappa\frac{1}{2}\{x_j, \vec{x}.\vec{p}\}, p_j=-\partial_j$. The conserve quantities of such a system are $\rho_0(=\omega_0A_0)$ and two components of Laplace-Runge-Lenz vector($\gamma $-deformed) given by
\begin{equation}
\rho_j=\frac{1}{2}\omega_0\{A_0, \epsilon_{jk}\pi_k\}+\omega_0\mu_0\frac{x_j}{r}\:\::j=1, 2
\end{equation}
where $ \epsilon_{jk}$ is the two dimensional Levi-Civita symbol.
Defining $\rho_{\pm}=\rho_1\pm i\rho_2$ we get the following relations for Higgs algebra
\begin{eqnarray}
[\rho_0, \rho_{\pm}]&=&\pm\omega_0\rho_{\pm}\nonumber\\
{[\rho_+, \rho_-]}&=&4\left[\left(\frac{\kappa}{8}-\mathcal{H}_{\kappa}\right)\omega_0\rho_0+\frac{\kappa}{\omega_0} \rho_0^3\right]\nonumber\\
\end{eqnarray}
 The Casimir of this algebra is found to be of the form
 \begin{equation}
 \mathcal{C}_{\rho}=\rho_-\rho_++\omega_0f(\rho_0; \omega_0)=\rho_+\rho_-+\omega_0f(\rho_0; -\omega_0)
 \end{equation}
 where
\begin{equation}
f(\rho_0; \omega_0)=2\left(\frac{\kappa}{8}-\mathcal{H}_{\kappa}\right)\rho_0(\rho_0-\omega_0)+\frac{\kappa}{\omega_0^2}\rho_0^2(\rho_0-\omega_0)^2
\end{equation}
On the other hand, choosing  $\lambda=\frac{\lambda_0}{\omega_0^{2(p+q)-r-2}}$ equation-24 can be  given as
\begin{eqnarray}
[H_{+}^{\gamma},H_{-}^{\gamma}]=4\left[\left(\frac{\lambda}{8}-\mathcal{U}_{\lambda}^{\gamma}\right)\omega_0 H_0^{\gamma}+\frac{\lambda}{\omega_0}(H_0^{\gamma})^3\right]
\end{eqnarray} 
with similar expression of Casimir $\mathcal{C_H^{\gamma}}=H_-^{\gamma}H_+^{\gamma}+\omega_0f(H_0^{\gamma};\omega_0)=H_+^{\gamma}H_-^{\gamma}+\omega_0f(H_0^{\gamma};-\omega_0)$.
The corresponding Hamiltonian can be given by
\begin{eqnarray}
\mathcal{U}_{\lambda}^{\gamma}=\frac{\lambda}{4\omega^2_0}(J_0^{\gamma}+K_0^{\gamma})^2+\lambda\left(\mathcal{C}+\frac{1}{8}\right)
\end{eqnarray}
It is to be noted that the non-hermitian operator $\mathcal{U}_{\lambda}^{\gamma}$ is a central element of the algebra as expected and it is defined for all allowed values of $\omega_0$ except $0$. 

Similar consideration for eqn-18 with $B_2=0=D_2$ leads to 
\begin{equation}
[Q_+^{\gamma}, Q_-^{\gamma}]=s_1^3[A_2(Q_0^{\gamma})^3+C_2Q_0^{\gamma}]
\end{equation}
Choosing $s_1^2=\frac{4\lambda_1}{A_2\omega_0}$ the following Hamiltonian
\begin{equation}
\mathcal{V}^{\gamma}_{\lambda_1}=\frac{\lambda_1}{8}-\frac{\lambda_1}{\omega_0^2}[\beta_2(\Lambda_0^{\gamma})^2+\beta_1\Lambda_0^{\gamma}+\beta_0]
\end{equation}
is possible. Here $\{\beta_j\vert j=1, 2, 3\}$ are rational functions of $\omega_0$ like the following
\begin{eqnarray}
\beta_2 &=&-\frac{2(1+3\omega_0)}{(1+\omega_0)(1+2\omega_0)}\nonumber\\
\beta_1 &=&\frac{3\omega_0^2+9\omega_0+2}{(1+\omega_0)(1+2\omega_0)}\nonumber\\
\beta_0 &=&\frac{12(3+\omega_0)\omega_0^2\mathcal{C_J^{\gamma}}-(1+\omega_0)^2(1+2\omega_0)-6\omega_0(1+\omega_0)}{12(1+\omega_0)(1+2\omega_0)}
\end{eqnarray}
$\mathcal{V}_{\lambda_1}^{\gamma}$ is defined for all allowed values of $\omega_0$ except $0$ and $-\frac{1}{2}$. Similar results can be expected by combining $su_{\gamma}(1, 1)$ and a boson algebra.

\subsection{Fusion of $su_{\gamma}(2)$ and $su_{\mu}(1, 1)$}
Introducing 
 \begin{eqnarray}
Y_0^{\gamma}&=&\frac{1}{2}(J_0^{\gamma}-Z_0^{\mu})\nonumber\\
Y_{\pm}^{\gamma,\mu}&=&s^{\prime} J_{\pm}^{\gamma}Z_{\mp}^{\mu}\nonumber\\
X_0^{\gamma,\mu}&=&\frac{1}{2}(J_0^{\gamma}+Z_0^{\mu})
\end{eqnarray}
we find 

\begin{eqnarray}
[Y_{0}^{\gamma},Y_{\pm}^{\gamma}]&=&\pm\frac{1}{2} (\omega_0(\gamma)+\omega_0(\mu))Y_{\pm}^{\gamma}\nonumber\\
{[Y_{+}^{\gamma}, Y_{-}^{\gamma}]}&=&(s^{\prime})^2[\Gamma_0+\Gamma_1Y_0^{\gamma}+\Gamma_2(Y_0^{\gamma})^2+\Gamma_3(Y_0^{\gamma})^3]
\end{eqnarray}
 where
\begin{eqnarray}
\Gamma_0 &=&\frac{2}{\omega_0^{2q-1}(\mu)\omega_0^{2p-1}(\gamma)}\left[{\mathcal{C_Z^{\mu}}}{\omega_0(\mu)}+{\mathcal{C_J^{\gamma}}}{\omega_0(\gamma)}\right]X_0^{\gamma, \mu}-2\Omega_{-}(X_0^{\gamma, \mu})^3\nonumber\\
\Gamma_1 &=&\frac{2}{\omega_0^{2p-1}(\gamma)\omega_0^{2q-1}(\mu)}\left[{\mathcal{C_K^{\mu}}}{\omega_0(\mu)}-{\mathcal{C_J^{\gamma}}}{\omega_0(\gamma)}\right]-2\Omega_{+}(X_0^{\gamma, \mu})^2\nonumber\\
\Gamma_2 &=&2\Omega_{-}X_0^{\gamma, \mu}\nonumber\\
\Gamma_3 &=&2\Omega_{+}
\end{eqnarray}

The possibility of Higg's algebra in the above case demands the conditions: $\gamma=\pm\mu$ and $\mathcal{C_Z^{\mu}}+\mathcal{C_J^{\gamma}}=0$. Hence the corresponding Higg's algebra is given by

\begin{eqnarray}
[Y_{0}^{\gamma},Y_{\pm}^{\gamma}]&=&\pm (\omega_0(\gamma))Y_{\pm}^{\gamma}\nonumber\\
{[Y_{+}^{\gamma},Y_{-}^{\gamma}]}&=& 4(s^{\prime})^2Y_{0}^{\gamma}(\gamma^{\prime}_1+\gamma^{\prime}_2(Y_0^{\gamma})^2)
\end{eqnarray}

where, $\gamma^{\prime}_1=-\frac{\mathcal{C^{\prime}}}{\omega_0^{2(p+q)-3}}-\frac{(X_0^{\gamma})^2}{\omega_0^{2(p+q)-1}}$ and $\gamma^{\prime}_2=\frac{1}{\omega_0^{2(p+q)-1}}$. Choosing $(s^{\prime})^2=\frac{\lambda^{\prime}}{\omega_0^r}$ one can write

\begin{equation}
[Y_{+}^{\gamma},Y_{-}^{\gamma}]=4\left[-\lambda^{\prime}\left(\frac{{\mathcal{C^{\prime}}}}{\omega_0^{2(p+q)-r-3}}+\frac{(X_0^{\gamma})^2}{\omega_0^{2(p+q)-r-1}}\right)Y_0^{\gamma}+\frac{\lambda^{\prime}}{\omega_0^{2(p+q)-r-1}}(Y_0^{\gamma})^3\right]
\end{equation}

Similar expressions for Casimir and multiboson Hamiltonian are possible.

\subsection{Deformed Hahn Algebra}
Given the Higgs algebra (as given in section-4.2)it possible to formulate Hahn algebra with the help of following operators
\begin{eqnarray}
\Theta_1^{\gamma}&=&H_1^{\gamma}+\frac{s}{\omega_0}(H_0^{\gamma})^2\nonumber\\
\Theta_2^{\gamma}&=&H_0^{\gamma}\nonumber\\
\Theta_3^{\gamma}&=&-iH_2^{\gamma}
\end{eqnarray}
Here, $H_1^{\gamma}=\frac{1}{2}(H_{+}^{\gamma}+H_{-}^{\gamma})$ and $H_2^{\gamma}=\frac{1}{2i}(H_{+}^{\gamma}-H_{-}^{\gamma})$. 
This leads to the commutation relations of a Hahn algebra in boson realization with cubic deformation as represented by the following commutation relations.
\begin{eqnarray}
[\Theta_1^{\gamma}, \Theta_2^{\gamma}]&=&\omega_0\Theta_3^{\gamma}\nonumber\\
{[\Theta_2^{\gamma}, \Theta_3^{\gamma}]}&=&s(\Theta_2^{\gamma})^2-\omega_0\Theta_1^{\gamma}\nonumber\\
{[\Theta_3^{\gamma}, \Theta_1^{\gamma}]}&=&2\left(\frac{s^2}{8\omega_0^2}+\mathcal{U}_{\lambda}^{\gamma}\right)\omega_0\Theta_2^{\gamma}+s\{\Theta_1^{\gamma}, \Theta_2^{\gamma}\}+\frac{2s^2\gamma^2}{\omega_0^3}(\Theta_2^{\gamma})^3
\end{eqnarray}

\section{Partial $\mathcal{PT}$-symmetry and spectrum of $J_0^{\gamma}$}
Considering $\mathcal{P}_j$ as an operator whose action is defined by the commutation relation $\{\mathcal{P}_j, a_j\}=0=\{\mathcal{P}_j, a^{\dagger}_j\}$ and $[\mathcal{P}_j, a_k]=0=[\mathcal{P}_j, a^{\dagger}_k]$ for $j\neq k$. Similar relations hold for $a_j^{\dagger}$ and $a_k^{\dagger}$. $P_j$ can be called the $j$-th partial parity operator. The time reversal operator $\mathcal{T}$ follows the rule $[\mathcal{T}, a_k]=0=[\mathcal{T}, a^{\dagger}_k]$ and it also changes $i=\sqrt{-1}$ to $-i$. Partial $\mathcal{PT}$ symmetry can be understood by the composite action of the operator $\Pi_T^j=\mathcal{P}_j\mathcal{T}$. This means $\{\Pi_T^j,a_j\}=0=\{\Pi_T^j,a^{\dagger}_j\}$ and $[\Pi_T^j,a_k]=0=[\Pi_T^j,a^{\dagger}_k]$.
\par It is to be noted that the operator $J_0$ has both global and partial $\mathcal{PT}$-symmetry where as $J_0^{\gamma}$ has only partial  $\mathcal{PT}$-symmetry i.e.; $[\Pi_T^j,J_0^{\gamma}]=0$. This holds for any central element of the subsequent formulation of quadratic and cubic algebras giving  $[\Pi_T^j,R_0^{\gamma}]=0=[\Pi_T^j,Q_0^{\gamma}]=[\Pi_T^j,H_0^{\gamma}]$. The related Casimirs obey the same type of relations i. e. ; $[\Pi^j_T, \mathcal{C_J^{\gamma}}]=0=[\Pi^j_T, \mathcal{C_R^{\gamma}}]=[\Pi^j_T, \mathcal{C_Q^{\gamma}}]=[\Pi^j_T, \mathcal{C_H^{\gamma}}]$
\par Considering $a_j^{\dagger}=\zeta_j$ and $a_j=\partial_{\zeta_j}$ the operators $J_0^{\gamma}$ and $J_{\pm}^{\gamma}$ take the form
\begin{eqnarray}
J_0^{\gamma}&=&\frac{1}{2}[\zeta_1\partial_{\zeta_1}-\zeta_2\partial_{\zeta_2}+i\gamma(\zeta_1\partial_{\zeta_2}+\zeta_2\partial_{\zeta_1})]\nonumber\\
J_{\pm}^{\gamma}&=&\frac{1}{2\omega_0^p}[(\zeta_1\partial_{ \zeta_2}+\zeta_2\partial_{ \zeta_1})-i\gamma(\zeta_1\partial_{\zeta_1}-\zeta_2\partial_{\zeta_2})]\pm\frac{1}{2\omega_0^{p-1}}(\zeta_1\partial_{\zeta_{2}}-\zeta_2\partial_{\zeta_{1}})
\end{eqnarray}
As the operator $J_0^{\gamma}$ leaves the homogeneous polynomial space of two indeterminates $(\zeta_1, \zeta_2)$ invariant one can find out the eigenvalues and corresponding eigenvectors in that space.  If such a space has degree of homogeneity $m$ and polynomial bases $\{f_k=\zeta_1^{m-k}\zeta_2^{k}\vert k=0\dots m\}$ the operator $J_0^{\gamma}$ has the following tri-diagonal representation.
\begin{eqnarray}
J_0^{\gamma}=\frac{1}{2}\mathcal{A}= \frac{1}{2}\left(\begin{array}{cccccccccc}
   m & i\gamma  &0 &0 &0 &\ldots&0 &0 \\
   im\gamma & m-2 &2i\gamma  &0 &0  &\ldots &0 &0 \\
   0  & i\gamma(m-1)  & m-4  &3i\gamma &0  &\ldots &0&0 \\
   \vdots &\vdots &\vdots &\vdots &\vdots & \ldots &-(m-2) &im\gamma\\
  0 &0 &0 &0 &0 &\ldots &i\gamma &-m
 \end{array} \right)
\end{eqnarray}
For $\gamma=0$ the eigenvalue of the operator $\mathcal{A}$ are integers across the diagonal from $-m$ to $+m$. For $\gamma\neq 0$ we shall make use of following results related to Gershgorin theorem.
\par 1.{\it Gershgorin disk theorem: If $\lambda$ is an eigenvalue of an $n\times n$ matrix $\mathcal{V}$}, 
{\it $\lambda\in G=\cup_{i=1}^n \mathcal{D}_i(v_{ii}, r_i)$}, {\it where, $\mathcal{D}_i$ is the i-th Gershgorin disc of radius $r_i=\sum_{j\neq i}\vert v_{ji}\vert$} {\it and centre at (Re$(v_{ii})$, Im$(v_{ii})$)}.
\par 2. {\it If a matrix has disjoint Gershgorin disks then each of them contains an eigenvalue. }
\par 3. {\it If a matrix with real diagonal elements has disjoint Gershgorin disks the eigenvalues are real}.

 As $\gamma$ takes up fractional values the eigenvalues are not necessarily integers but they are evenly distributed on either side of the imaginary axis and confined in the set (according to Gershgorin disk theorem) $S_{\mathcal{A}}=\bigcup_{k=0}^m\mathcal{D}_k(m-2k; m\gamma)$. Here $\mathcal{D}_k(c, r)$ is  the {\it k-th Gershgorin disk} of radius $m\gamma$ and center $(m-2k, 0)$. As the centres of the disks (diagonal terms) are equidistant on the real axis and all the disks have same radii $(m\gamma)$ it  can be concluded that for a given $m$ there exists a range $0<\gamma<1/m$ of $\gamma$ values so that $m+1$ number of distinct real eigenvalues are possible corresponding to $m+1$ number of possible disjoint Gershgorin disks each of them containing an eigenvalue. As the value of $\gamma$ increases from $1/m$ to $1$ the overlap region between two consecutive disks increases. In this range the reality and distinctness of eigenvalues cannot be ensured as such.  The increase in radii of the disks is indicative of the fact that if the eigenvalues are monotonic functions of $\gamma$ and still real they are either moving towards the origin or towards infinity and hence distancing themselves more and more from the centre of the respective Gershgorin disks. One can expect the possibility of maximum overlap for $\gamma=1$ corresponding to the largest possible Gershgorin radius. The following exact estimation of eigenvalues shows that they remains to be real for $\frac{1}{m}\leq\gamma<1$ and they indeed gravitate towards the origin as $\gamma$ increases.
\par In order to obtain eigenvalues and eigenvectors of $\mathcal{A}$ we use the following theorem\cite{sandry13}:
\par{\it Given a tri-diagonal matrix }
\begin{eqnarray}
\mathcal{M}= \left(\begin{array}{cccccccccc}
   b_0 & d_0  &0 &0&0 &\ldots &0&0 \\
   c_0  & b_1  &d_1 &0  &0  &\ldots &0 &0 \\
    0  & c_1  &b_2 &d_2  &0  &\ldots &0 &0 \\
   \vdots &\vdots &\vdots &\vdots & \vdots & \ldots & b_{l-2} & d_{l-2}\\
   0&0 &0&0 &0&\ldots &c_{l-1}&b_{l-1}
 \end{array} \right)
\end{eqnarray}

{\it with} $d_i\neq 0 \forall i$ {\it let us consider a polynomial} $P_n(x)$ {\it that follows the recursion relation}
\begin{equation}
P_{n+1}(x)=\frac{1}{d_n}[(x-b_n)P_n(x)-c_{n-1}P_{n-1}(x)].
\end{equation}
{\it If} $P_{-1}(x)=0$ {\it and} $P_0(x)=1$ {\it the eigenvalues are given by the zeros of the polynomial} $P_l(x)$ {\it and eigen-vector corresponding to} $j$-{\it th eigenvalue $x_j$ is given by the the vector }

\begin{eqnarray}
 \left(\begin{array}{c}
  P_0(x_j)  \\
   P_1(x_j) \\
   \vdots\\
   P_{l-2}(x_j)\\
   P_{l-1}(x_j)  
    \end{array} \right)
\end{eqnarray}

\par For $m=2k$ and $m=2k+1$ the polynomial equation $P_{m+1}(x)=0$ can be given by
\begin{eqnarray}
x\prod_{j=1}^k[x^2-(2j\omega_0)^2]=0\nonumber\\
\prod_{k=1}^k[x^2-(2j+1)^2\omega_0^2]=0
\end{eqnarray}
respectively.
This leads to the following set of eigenvalues of the operator $J_0^{\gamma}$

\begin{eqnarray}
S_{n=2k}^{\omega_0}&=&\{-k\omega_0,-(k-1)\omega_0\dots ,0, \dots, (k-1)\omega_0, k\omega_0 \}\nonumber\\
S_{n=2k+1}^{\omega_0}&=&\left\{-\frac{2k+1}{2}\omega_0,-\frac{2k-1}{2}\omega_0\dots -\frac{1}{2}\omega_0 , \frac{1}{2}\omega_0 \dots, \frac{2k-1}{2}\omega_0, \frac{2k+1}{2}\omega_0 \right\}
\end{eqnarray}
and the corresponding eigenfunctions are of the form
\begin{eqnarray}
\psi_{(\pm l\omega_0)}(\zeta_1, \zeta_2)&=&F^{\pm}_{2l}(\omega_0)\zeta_1^{2l}+iF^{\pm}_{2l-1}(\omega_0)\zeta_1^{2l-1}\zeta_2+\dots +F^{\pm}_0(\omega_0)\zeta_2^{2l}\nonumber\\
\psi_{\left(\pm\frac{2l+1}{2}\omega_0\right)}(\zeta_1, \zeta_2)&=&G^{\pm}_{2l+1}(\omega_0)\zeta_1^{2l+1}+iG^{\pm}_{2l}(\omega_0)\zeta_1^{2l}\zeta_2+\dots+i G^{\pm}_0 (\omega_0)\zeta_2^{2l}
\end{eqnarray}
Where $\{F_j, G_j\}$ are real functions of $\omega_0$.
It is easy to verify that $\psi_{(l\omega_0)}$ is partial $\mathcal{PT}$-symmetric in either of the variables while $\psi_{\left(\frac{(2l+1)\omega_0}{2}\right)}$ is showing a kind of $breaking\:\: of \:\: partial \:\:\mathcal{PT}-symmetry $ in the sense that if it shows partial  $\mathcal{PT}$-symmetry  in one variable the other variable does not conform to such a symmetry. Thus we have partial  $\mathcal{PT}$-symmetry conforming state alternated by partial  $\mathcal{PT}$-symmetry breaking state.

 \par For example considering the space of homogeneous symmetric polynomial of degree $2$ spanned by $\{\zeta_1^2,  \zeta_1\zeta_2, \zeta_2^2\}$ the eigenfunctions of $J_0^{\gamma}$ corresponding to eigenvalues $\{\pm\omega, 0 \}$ are given by
\begin{eqnarray}
\psi_{(\pm\omega_0)}(\zeta_1, \zeta_2)&=&-\frac{1\mp\omega_0}{1\pm\omega_0}\zeta_1^2-2i\left(\frac{1\pm\omega_0}{1\mp\omega_0}\right)^{\frac{1}{2}}\zeta_1\zeta_2+\zeta_2^2\nonumber\\
\psi_{(0)}(\zeta_1, \zeta_2)&=&-\zeta_1^2-\frac{2i}{\sqrt{1-\omega_0^2}}\zeta_1\zeta_2+\zeta_2^2\nonumber\\
\end{eqnarray}
Similarly for space of polynomial of degree $3$ spanned by $\{\zeta_1^3,  \zeta_1^2\zeta_2, \zeta_1\zeta_2^2, \zeta_2^3\}$ one can write eigenfunctions corresponding to the eigenvalues $\{\pm\frac{3}{2}\omega, \pm\frac{1}{2}\omega\}$ as given below
\begin{eqnarray}
\psi_{(\pm\frac{3}{2}\omega_0)}(\zeta_1, \zeta_2)&=&\left(\frac{1\pm\omega_0}{1\mp\omega_0}\right)^{\frac{3}{2}}\zeta_1^3+3i\frac{1\pm\omega_0}{1\mp\omega_0}\zeta_1^2\zeta_2-3\left(\frac{1\pm\omega_0}{1\mp\omega_0}\right)^{\frac{1}{2}}\zeta_1\zeta_2^2-i\zeta_2^3\nonumber\\
\psi_{(\pm\frac{1}{2}\omega_0)}(\zeta_1, \zeta_2)&=&\left(\frac{1\pm\omega_0}{1\mp\omega_0}\right)^{\frac{1}{2}}\zeta_1^3-i\left(1\pm\frac{2}{1\pm\omega_0}\right)\zeta_1^2\zeta_2-\frac{3\pm\omega_0}{(1-\omega_0^2)^{1/2}}\zeta_1\zeta_2^2-i\zeta_2^3
\end{eqnarray}
Obviously $\{\psi_{(\pm\omega_0, 0)}\}$ are partial  $\mathcal{PT}$-symmetry conforming states while $\{\psi_{(\pm3\omega_0/2, \pm\omega_0/2)}\}$ partial  $\mathcal{PT}$-symmetry breaking states. It is to be noticed that the eigenfunctions are non-trivially modified in the sense letting $\gamma\rightarrow 0$ i. e.; $\omega_0\rightarrow1$ eigenfunctions corresponding to the hermitian case cannot be retrieved.

\section{Conclusion}
The main objective of our present discussion is to construct polynomial algebras starting from certain kind of deformed Lie algebras and their dependence on the deformation parameters ($\gamma$ or $\mu$ or $\omega_0$) as well as their exponents ($p, q, r$). Such deformation can be meaningful in the sense they give rise to  non-hermitian operators and Hamiltonians which are related to one of the current trends in Quantum Mechanics. The operators we obtained through boson realization may find their relevance in many-particle systems. The notion of partial  $\mathcal{PT}$ symmetry as it is currently understood\cite{beygi15} in a typically many particle context has been used in case of one such operator ($J_0^{\gamma}$) whose eigenvalues are found to be real for any allowed value of the deformation parameter. The eigenfunctions of the same operator are found to be showing symmetry-obeying and symmetry-breaking states. The present attempt of boson realization seems to motivate the widening  of non-hermitian regime in various many particle theories.

\vspace{.5cm}

\noindent {\bf \Large Acknowledgement}
\vspace{.1cm}

A.C. wishes to thank his colleague Dr. Baisakhi Mal for her valuable assistance in preparing the latex version.

\end{document}